\RequirePackage{lineno} 
\documentclass[aps,preprint,showpacs,superscriptaddress,groupedaddress,nofootinbib]{revtex4}  

\usepackage{graphicx}  
\usepackage{dcolumn}   
\usepackage{bm}        
\usepackage{longtable}
\usepackage{siunitx}

\newcommand{ \be }{\begin{eqnarray}}
\newcommand{ \ee }{\end{eqnarray}}

\usepackage{lineno}

\begin{document}

\title{Particle rapidity distribution in proton-nucleus collisions using the proton-contributor reference frame}

\author{Gin\'es Mart\'{\i}nez-Garc\'{\i}a }
\affiliation{Subatech (CNRS/IN2P3 - Ecole des Mines de Nantes - Universit\'e de Nantes), Nantes, France}

\date{\today}

\modulolinenumbers[5]

\begin{abstract}
I define the proton-contributor reference frame in proton nucleus (p--A) collisions 
as the center of mass of the system formed by the proton and the 
participant nucleons of the nucleus. Assuming that the rapidity distribution of produced 
particles is symmetric in the proton-contributor reference frame, several 
measurements in p--Pb collisions at $\sqrt{s_{\rm NN}} = 5.02~{\rm TeV}$ can 
be described qualitatively.
These include rapidity distributions of charged particles, $J/\psi$ and Z bosons. 
\end{abstract}


\maketitle

\section{Introduction}

In proton-proton (pp) and nucleus-nucleus (A--A) collisions, the colliding system defines 
unambiguously the reference frame in which the rapidity distribution 
of particles produced in the collisions must be symmetric.
In proton-nucleus (p--A) collisions, the situation is more complex since this rapidity 
distribution is not expected to be necessarily symmetric.
Therefore we are used to consider the nucleon-nucleon center-of-mass frame,
as for pp and A--A collisions. 
This is justified for high-energy p--A collisions which can 
be seen as a multiple interaction of parton pairs, one parton of the pair belonging to the 
proton and the other belonging to one of the nucleons of the nucleus.
I propose hereafter an alternative reference frame, called the 
{\it proton-contributor} reference frame.
The main assumption is that particles are produced in p--A collisions with 
a rapidity distribution identical to that in pp collisions, but shifted by 
the rapidity gap between the proton-nucleon and the proton-contributor 
reference frame. Under this simple hypothesis the centrality dependence 
of charged particle pseudo-rapidity distribution measured by the ATLAS 
collaboration, the suppression (enhancement) of the $J/\psi$ 
at forward (backward) rapidity measured by the ALICE 
collaboration, and  the Z boson backward-to-forward ratio measured by 
the CMS collaboration, are qualitatively understood.

\section{Time scale of the p--A collision}
\begin{figure}{\centering 
\resizebox*{0.90\columnwidth}{!}{\includegraphics{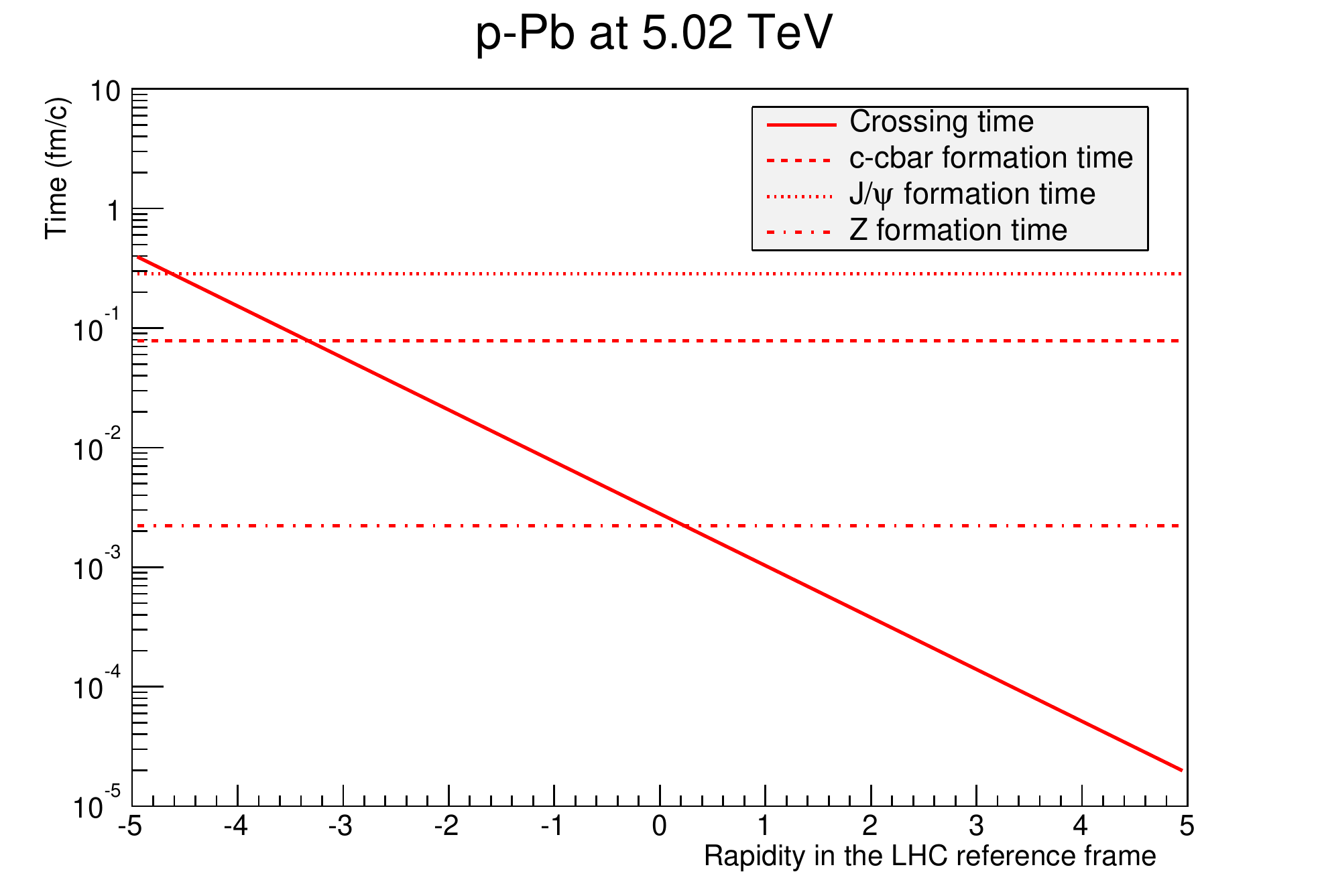}}
\par}
\caption{\label{fig:fig1} Formation times for J/$\psi$, $c\bar{c}$ pair and Z particles
and the crossing time in p--Pb collisions at 5.02 TeV, as a function 
of the probe rapidity in the LHC reference frame.}
\end{figure}

Many of the numerous theoretical models aiming at describing heavy-ion or 
proton-nucleus collisions at RHIC and LHC energies assume that, 
at sufficiently low transferred momentum, the 
interaction takes place coherently with all the partons of the nuclei or nucleus.
Such a coherent interaction will occur when the crossing time of the 
projectile and the target is smaller than the formation time of a given 
process.
Let us consider the production of a probe involving a momentum transfer 
$Q$. The formation time of the probe can be estimated as
\begin{equation}
\tau_f \approx Q^{-1}.
\end{equation}
In a proton-nucleus collision, the crossing-time  in the reference of the 
probe (centre of mass frame of the parton-parton interaction in 2$\rightarrow$1 processes) can be estimated as  
\begin{equation}
\tau_c \approx R/\gamma_R
\end{equation}
where R is the radius of the nucleus and $\gamma_R$ is the nucleus Lorentz factor 
in the probe reference frame.  $\gamma_R$  can be expressed as 
\begin{equation}
\gamma_R = \cosh{(y-y_A)}
\end{equation}
where $y_A$ is the rapidity of the nucleus and $y$ the rapidity of the probe. 
In p--Pb collisions at 5.02 TeV and for $y=0$, the crossing time $\tau_c$ is smaller than $\tau_f$ 
for $Q \lesssim70$ GeV.  Fig.\ref{fig:fig1} shows the formation times 
for J/$\psi$, $c\bar{c}$ pair and Z particles, compared to the crossing time 
in p--Pb collisions at 5.02 TeV as a function of the probe rapidity in the 
LHC reference frame. 
	
\section{The proton-contributor reference frame in p--A collisions}

We are used to consider the proton-nucleon reference frame to study 
the production of a probe in p--A collisions, namely when $\tau_c \gg \tau_f$. 
Indeed, the probe can be viewed as produced in a single collision of the 
proton with one of the nucleons of the nucleus. 
However, we have seen in the previous section that most of the time $\tau_c \leq \tau_f$ at LHC energies.
Therefore, the whole volume of the nucleus crossed by the proton (a cylinder of 
about $\sqrt{\sigma_{\rm NN}/\pi}\approx$1.5 fm radius, that I am calling \emph{contributor} in this paper) 
will coherently contribute to the production of the probe. 
In addition, the Bjorken-$x$ ($x_{Bj}$) values of the partons involved in 
the hard collision are small ($x_{Bj}\leq 2 \cdot 10 ^{-2}$ 
for Q$\leq 100$ GeV/$c$ at y=0 and $\sqrt{s}$=5 TeV). 
One could wonder whether the belonging of a small $x_{Bj}$ parton 
to a given nucleon is not \emph{blurred} by the presence of other 
nucleons contributing to the collision. This is the \emph{main} physics motivation\footnote{I agree that this physics motivation is weak, as my colleague St\'ehane Peign\'e already told me. Indeed, the main motivation to formulate the proton-contributor hypothesis is the successful explanation of several different phenomenological observations in p--A collisions at LHC energy as it is discussed in the present draft. 
} to make 
the extreme hypothesis that particles in p--A collisions at the LHC are 
produced with a rapidity differential cross section which is symmetric 
in the proton-contributor reference frame with a similar shape as in pp collisions:
\begin{equation}
\label{Eq:hypo}
\frac{{\rm d}N^{\rm probe}_{\rm pA(Ap)}}{{\rm d}y} \Bigl( y \Bigr) = {\cal N} \frac{{\rm d}N^{\rm probe}_{\rm pp}}{{\rm d}y} \Bigl( y -(+) \Delta y_{\rm pN-pC} \Bigr) 
\end{equation}
where ${\rm d}N/{\rm d}y$ are the probe yields, ${\cal N}$ is a normalisation parameter, 
and $\Delta y_{\rm pN-pC}$ is the rapidity gap between the proton-nucleon 
and proton-contributor frames, which is defined below.

The mass and momentum of the contributor can be obtained using the Glauber model:
\begin{equation}
m_{\rm C} = N_{\rm coll}(b) \times m_{\rm N}
\end{equation}
\begin{equation}
P_{\rm C} = N_{\rm coll}(b) \times  P_{\rm Pb}
\end{equation}
where $m_{\rm N}$ is the mass of the nucleon (here 931 MeV/c$^2$), $P_{\rm Pb}$ is the momentum per nucleon of the Pb LHC beam and $b$ the impact parameter.

The total energy of the proton-contributor system is given by:
\begin{equation}
E_{\rm pC} = \sqrt{P_p^2 + m_p^2} + \sqrt{P_{\rm C}^2 + m_{\rm C}^2}
\end{equation}
where $P_{\rm p}$ is the momentum of the LHC proton beam, and $m_{\rm p}$ is its 
mass (938 MeV/c$^2$). The total momentum (positive value in the 
direction of the proton beam) is
\begin{equation}
P_{\rm pC} = P_{\rm p} - P_{\rm C}.
\end{equation}
Finally, the rapidity of the proton-contributor in the laboratory frame is given by
\begin{equation}
y_{\rm pC} = \tanh^{-1}{(p_{\rm pC} /E_{\rm pC})}
\end{equation}

Assuming $N_{\rm coll}$=1, the rapidity  becomes $y_{\rm pC}$= 0.465, which is equal 
to the rapidity of the proton-nucleon frame.  In minimum bias p--Pb collisions 
at 5.02 TeV, the average number of collisions in the nucleonic tube is $\approx$6, 
therefore the rapidity of the proton-contributor system is $y_{pC}$=-0.430. 
The rapidity gap between the proton-nucleon and proton-contributor is close 
to one unit of rapidity, $\Delta y _{\rm pN-pC}$=0.896. For the most central p--A collisions 
($N_{\rm coll}=17$), the rapidity is $y_{\rm pC}$=-0.951.

\section{Centrality dependence of the charged particle pseudorapidity distribution 
in p-Pb collisions at  5.02 TeV}
Let us assume that the charged particle rapidity density ${\rm d}N^{ch}_{pp}/{\rm d}y$ 
exhibits a Gaussian 
shape with a width $\sigma$ of 3.2 rapidity units\footnote{The  $\sigma$ 
parameter of the charged particle rapidity Gaussian distribution has been chosen arbitrarily to reproduce with ATLAS measurement of the charged particle pseudo-rapidity distribution ratios \cite{ATLASQM2014:2014}.}.  
The charged particle rapidity density as a function of the centrality in p--A collisions $dN^{ch}_{pA}/dy$ can
be obtained applying Eq.\ref{Eq:hypo}.

The pseudo-rapidity density is 
then given by:
\begin{equation}
\label{eq:eqdNchdeta}
\frac{{\rm d}N_{ch}}{{\rm d}\eta} = \frac{{\rm d}N_{ch}}{{\rm d}y} \times \frac{{\rm d}y}{{\rm d}\eta}
\end{equation}
where
\begin{equation}
\theta = 2 \cdot \arctan{(e^{-\eta})}
\end{equation}
\begin{equation}
m_{\rm T} = \sqrt{p_{\rm T}^2+m^2}
\end{equation}
\begin{equation}
p_{\rm z} = \frac{p_{\rm T}}{\tan{\theta}} 
\end{equation}
and
\begin{equation}
y = \sinh^{-1} {\Biggl( \frac{p_{\rm z}}{m_{\rm T}} \Biggr)}
\end{equation}

\begin{figure}{\centering 
\resizebox*{0.85\columnwidth}{!}{\includegraphics{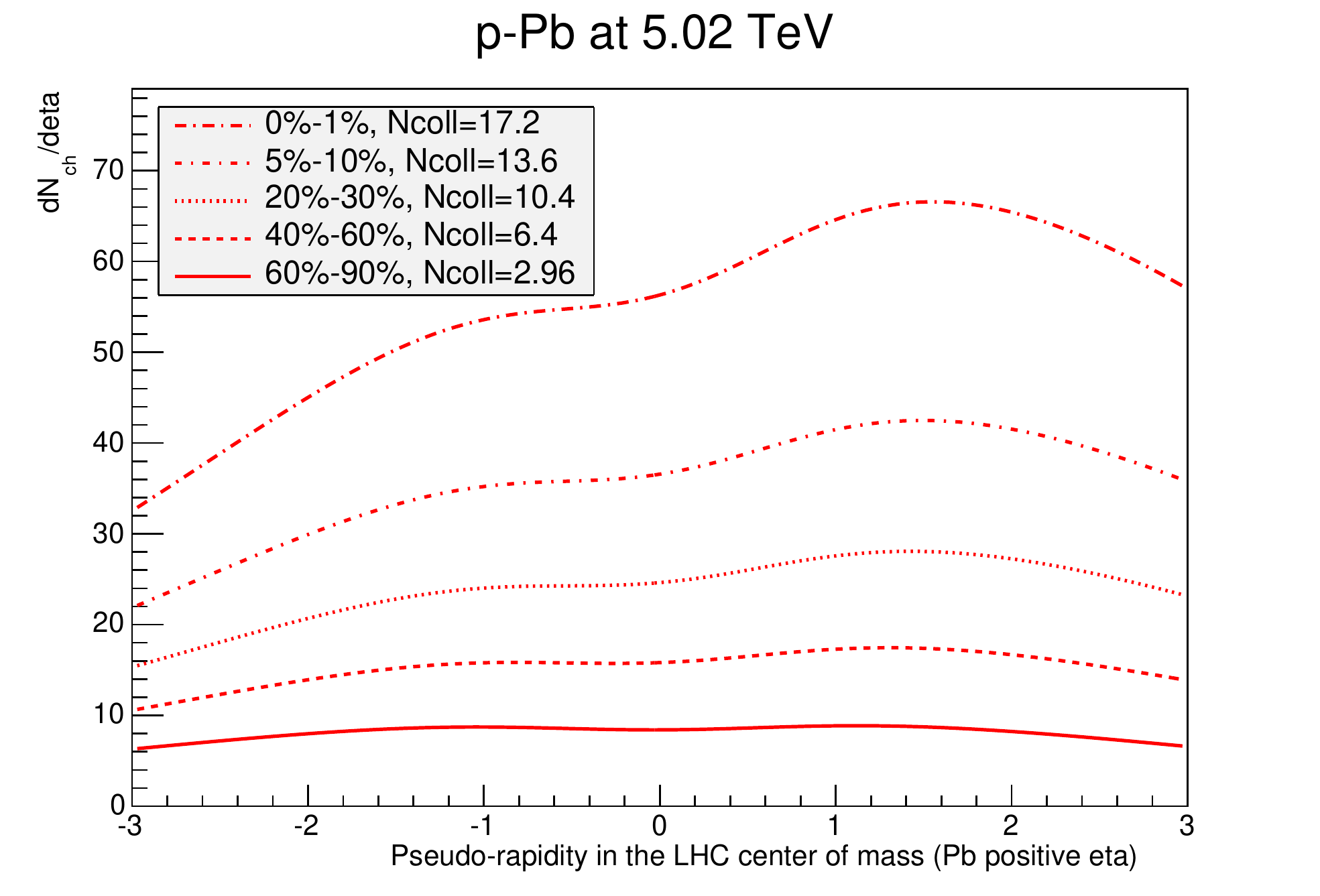}}
\resizebox*{0.85\columnwidth}{!}{\includegraphics{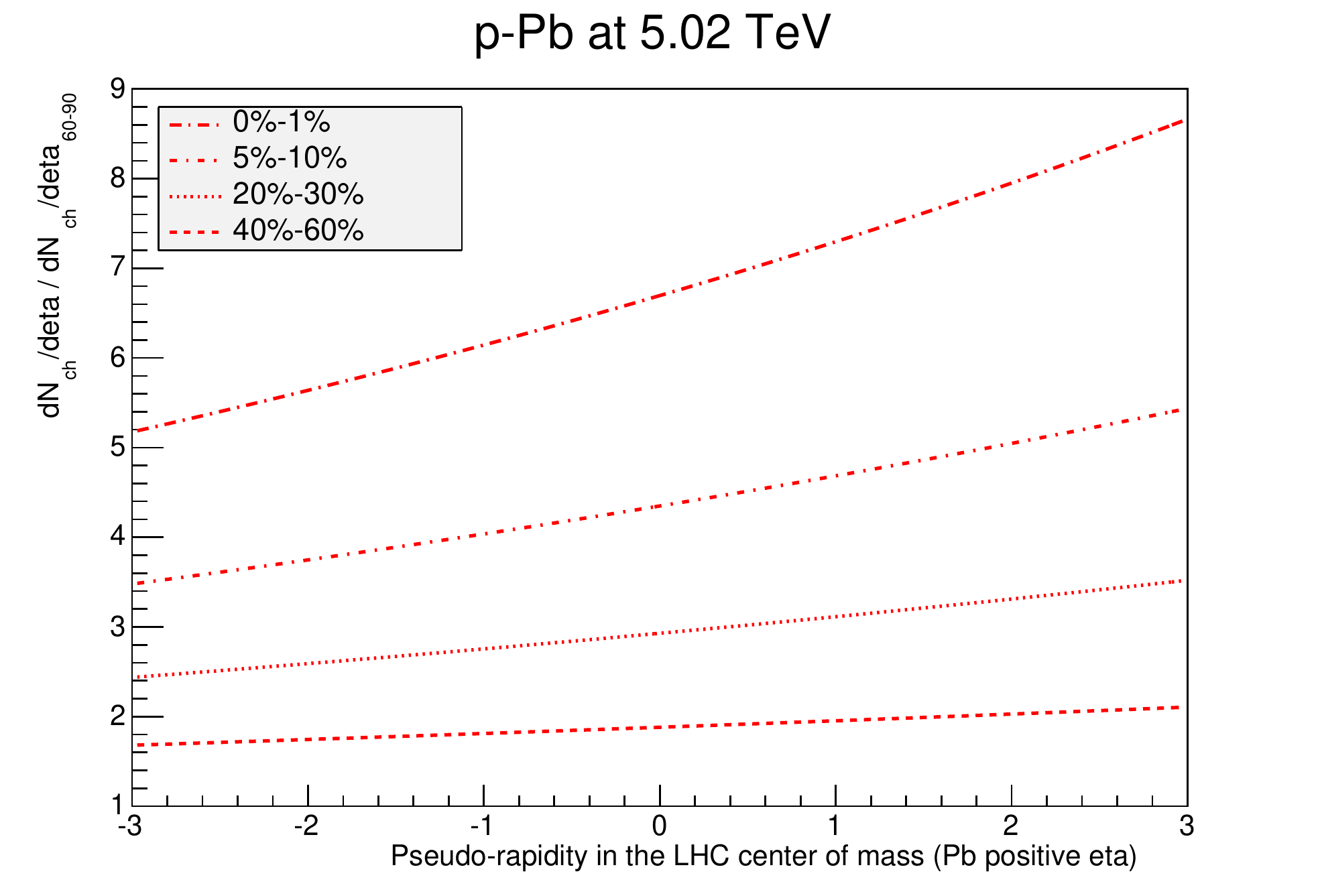}}
\par}
\caption{\label{fig:fig2} Top ${\rm d}N_{ch}/{\rm d}\eta$
in different centrality classes, assuming a  charged particle rapidity distribution 
with a Gaussian shape ($\sigma=3.2$) in the proton-contributor reference frame. 
Bottom: Ratios of ${\rm d}N_{ch}/{\rm d}\eta$ distributions in different centrality 
classes, with respect to ${\rm d}N_{ch}/{\rm d}\eta$ distribution in the peripheral (60\%-90\%) centrality interval. }
\end{figure}

The Jacobian depends on the particle mass and transverse momentum. 
For simplicity, I have assumed a mean charged particle mass of 450 MeV/$c^2$ 
and a mean transverse momentum of 700 MeV/$c$ \footnote{These values represent a first 
approximation, which could be improved considering realistic particle ratios and 
realistic particle transverse momentum distributions.}.  
The charged particle pseudo-rapidity densities as obtained from Eq.\ref{eq:eqdNchdeta} 
for a Gaussian rapidity distribution, are plotted in Fig.\ref{fig:fig2} top. 
The ${\rm d}N^{ch}_{\rm pA}/{\rm d}y$  distributions  are 
normalized  (${\cal N}$ parameter) to the charged particle pseudo-rapidity density at $\eta = 0$ 
measured by the ATLAS collaboration \cite{ATLASQM2014:2014}.  
 We observe that the shape of ${\rm d}N_{ch}/{\rm d}\eta$ becomes progressively 
 more asymmetric in the Pb-going direction, in accordance with the increase of the 
 contributor size and therefore the increase of the rapidity shift between proton-contributor frame
 in more central collisions and the proton-contributor frame in the peripheral 60\%-90\% centrality bin.
 In Fig.\ref{fig:fig2} bottom,  the ratio of the pseudo-rapidity densities  with respect 
to that in 60\%-90\% centrality bin is presented. 
The double peak structure present in the distributions in Fig.\ref{fig:fig2} top  
disappears in the ratios. The ratios are observed to grow
nearly linearly with pseudo-rapidity, and the slope increases from peripheral 
to central collisions. 

The ATLAS collaboration presented at the Quark Matter Conference in Darmstadt,
the charged particle pseudo-rapidity distribution in p--Pb collisions at 5.02 TeV as a function 
of the collision centrality \cite{ATLASQM2014:2014, DebbeVelasco:2014}.
A linear dependence of the charged particle pseudo-rapidity ratios 
with a slope increasing from peripheral to central p--Pb collisions, is observed (see Fig.8 in \cite{ATLASQM2014:2014}),
qualitatively agreeing the predictions presented in Fig.\ref{fig:fig2} bottom, obtained with the hypothesis in Eq.\ref{Eq:hypo}.

\section{Centrality dependence of the $J/\psi$ production in p--Pb collisions}

\begin{figure}{\centering 
\resizebox*{0.90\columnwidth}{!}{\includegraphics{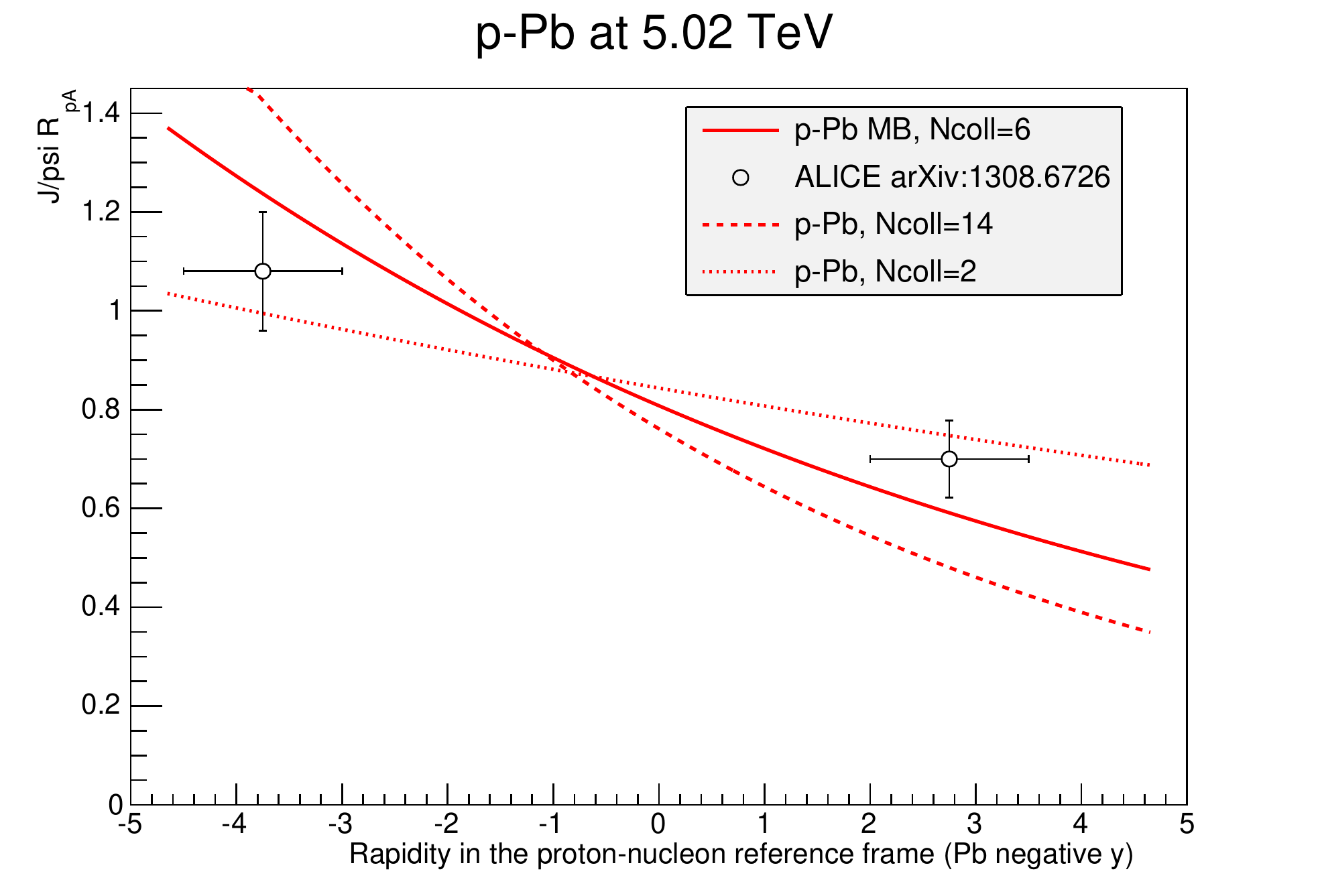}}
\par}
\caption{\label{fig:fig3}  Nuclear modification factor of J/$\psi$ in p--Pb 
collisions at 5.02 TeV assuming that the rapidity distribution 
shape is identical to that in pp, but shifted by the rapidity gap between 
the proton-nucleon and the proton-contributor frames. 
A suppression factor of 0.85 has been considered to mimic shadowing effect. 
The R$_{\rm pA}$ for $N_{\rm coll}=6$ is compared with the 
measurement performed by the ALICE collaboration.  
The R$_{\rm pA}$ for $N_{\rm coll}=2$ (peripheral p--Pb collision) and $N_{\rm coll}=14$ (central p--Pb collisions) are also shown.}
\end{figure}

Recently, the ALICE collaboration has published results on $J/\psi$ production 
and nuclear effects in p--Pb collisions at $\sqrt{s_{NN}}$ = 5.02 TeV \cite{Abelev:2013yxa}. 
Let us assume a J/$\psi$ rapidity distribution according to the 
phenomenological parameterization introduced in \cite{Bossu:2011qe}:
\begin{equation}
 \frac{{\rm d}\sigma}{{\rm d}y} \, \,/  \,\, \frac{{\rm d}\sigma}{{\rm d}y} \Bigr|_{y=0} =  e^{ -(y / y_{\rm max})^2/2\sigma_y^2}
\end{equation}
where $y_{\rm max}=\ln{(\sqrt{s/}m_{{\rm J}/\psi} ) }$ and $\sigma$=0.38 for pp collisions.
In p--A collisions, I am assuming that the rapidity distribution is shifted by the rapidity 
gap between the proton-nucleon and the proton-contributor frames following Eq.\ref{Eq:hypo}. 
The J/$\psi$ nuclear modification factor is then obtained from the 
ratio of the two Gaussian distributions. 
The normalization factor ${\cal N}$ is determined assuming binary scaling 
and an additional  \emph{shadowing-like} factor of 0.85\footnote{The only motivation of this 
shadowing-like factor is to fit the experimental data. Note 
that shadowing is expected to be almost constant as a function 
of rapidity and the 0.85 shadowing factor agrees very well with the predictions \cite{Abelev:2013yxa}.}. 
This is shown in Fig.3 together with the ALICE measurements. 
As it can be seen, the simple assumption of a $J/\psi$ rapidity 
distribution in p--A collisions shifted with respected to that in pp collisions 
allows to describe the observed $J/\psi$ suppression (enhancement) at 
forward (backward) rapidity.
The rapidity gap between the proton-nucleon and proton-contributor 
frames, would explain the observed $J/\psi$ suppression at forward 
rapidity and the enhancement at backward rapidities. 
As shown in Fig.\ref{fig:fig3}, the previous pattern is enhanced in 
central p--Pb collisions since the contributor size increases, thus 
increasing the rapidity gap. 
This is in qualitative agreement with results from the ALICE collaboration showing 
that the J/$\psi$ nuclear modification factor decreases with centrality at 
forward rapidity, while it increases with centrality at forward rapidity \cite{MartinBlanco:2014, Lakomov:2014}.

\section{Production of Z in p--A collisions at 5.02 TeV}

Recently, the ATLAS and CMS collaboration have reported the 
measurement of the Z  differential cross section in p--Pb collisions 
at 5.02 TeV \cite{ZATLASQM2014:2014, ZCMSQM2014:2014, Cole:2014, Granier:2014}.
ATLAS claimed that the rapidity differential cross section 
shows a significant asymmetry compared to the simple model
based on binary scaling with respect to nucleon-nucleon collisions. 
Indeed, a relative excess in the Z  differential cross section 
is seen in the backward (Pb-going) part of the rapidity distribution \cite{ZATLASQM2014:2014}. 
CMS interpreted such an asymmetry as a consequence of the 
modification of the parton distribution functions (PDF) in the 
nucleus and claimed that this measurement is providing new data 
points in a previously unexplored region of phase space for 
constraining nuclear PDF fits \cite{ZCMSQM2014:2014}.
Furthermore, ATLAS  observed a more pronounced asymmetry 
in central events, while the asymmetry is apparently absent in peripheral events.

\begin{figure}{\centering 
\resizebox*{0.85\columnwidth}{!}{\includegraphics{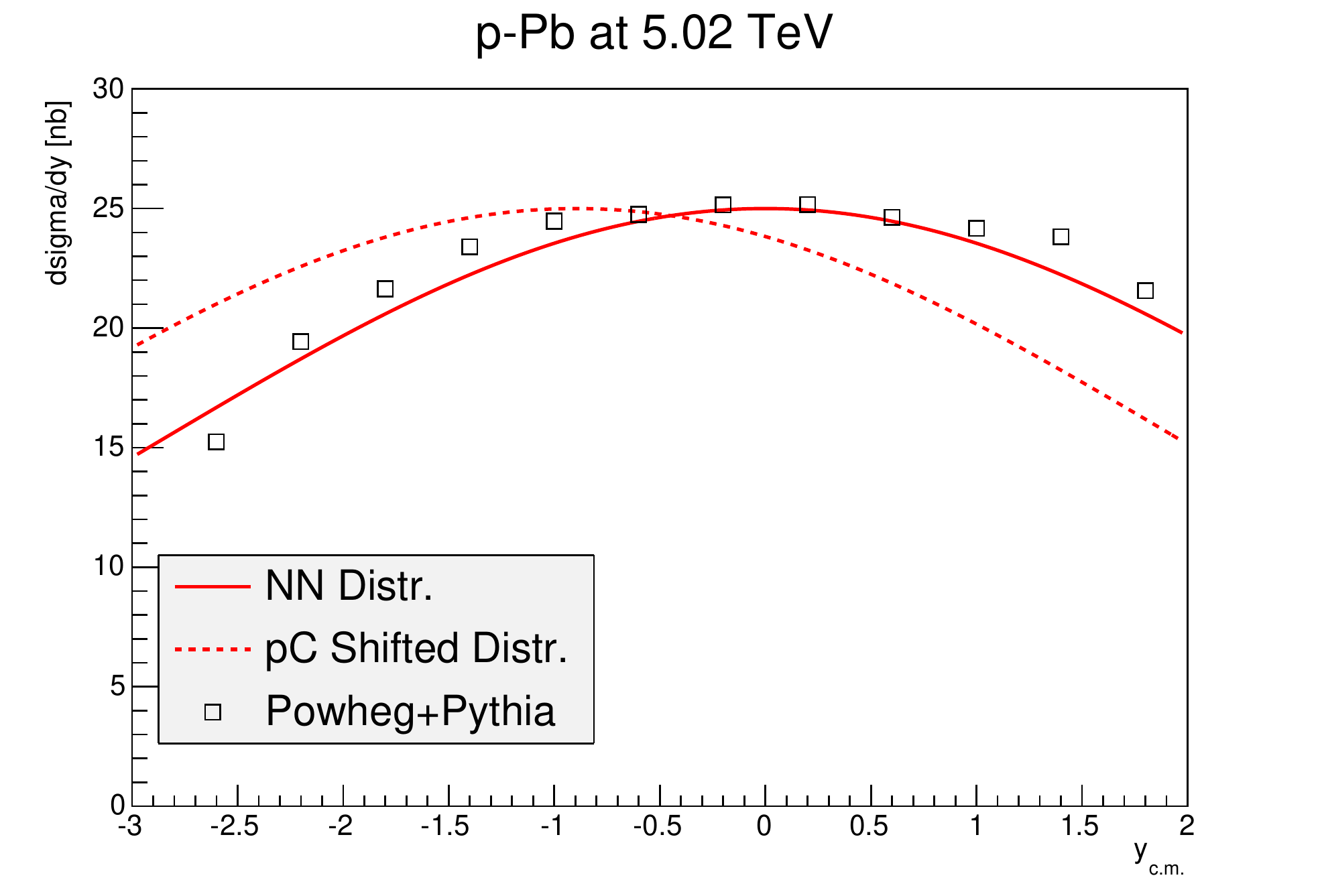}}
\resizebox*{0.85\columnwidth}{!}{\includegraphics{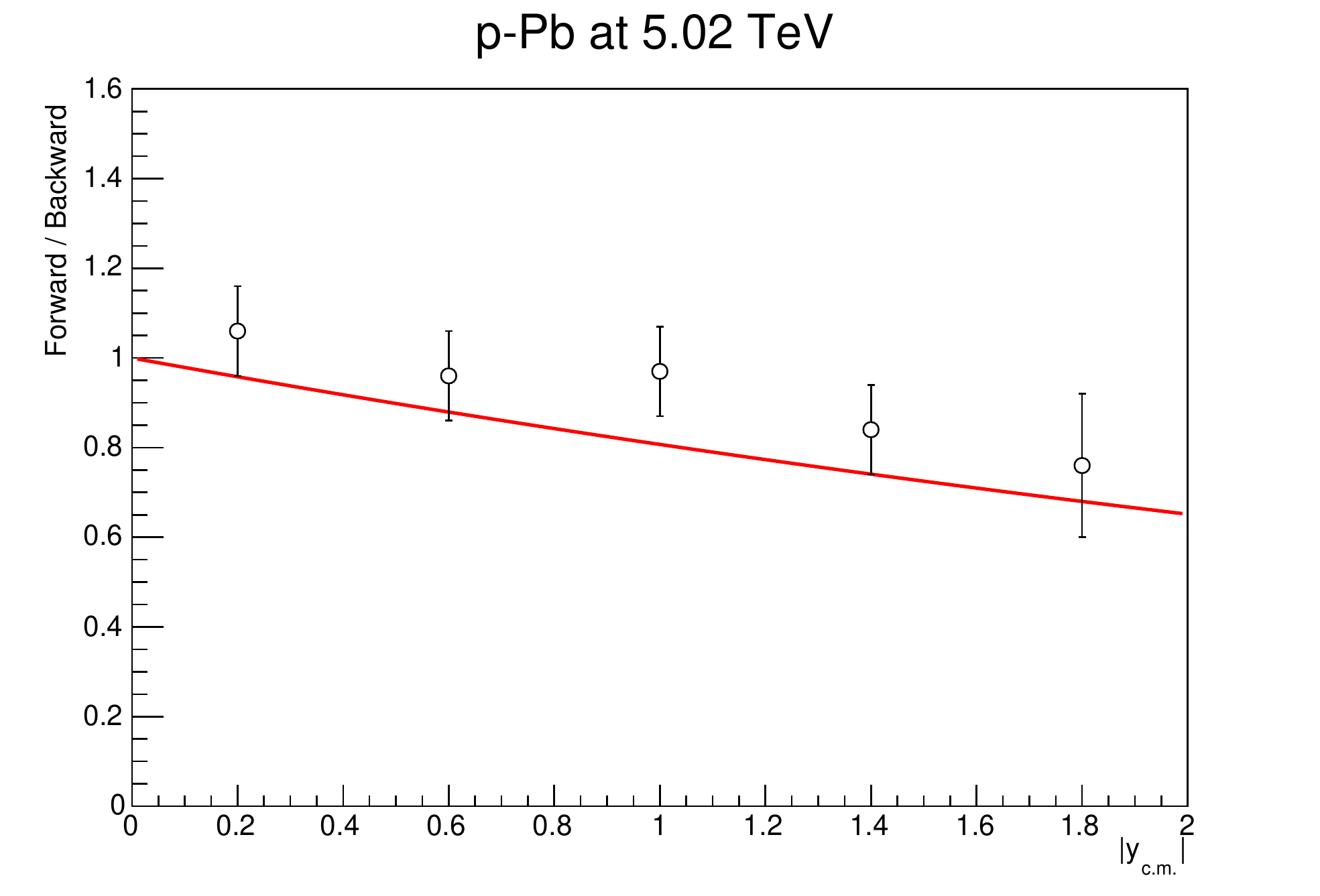}}
\par}
\caption{\label{fig:fig4} Top: Z  rapidity distributions in the 
proton-nucleon centre of mass frame. The solid curve models the  
Powheg-Pythia predictions (open symbols)  presented in Fig.2 of 
\cite{Granier:2014b, ZCMSQM2014:2014}. 
The dashed curve is the same distribution but is centred at the 
proton-contributor rapidity. Bottom: The ratio of the dashed and solid 
curve allows to estimate the backward-to-forward ratio of the Z 
production in the proton-nucleon centre of mass. 
Open circles represent the  backward-to-forward ratio measured by CMS, 
extracted from the pdf file, Fig.3 of  \cite{ZCMSQM2014:2014}. }
\end{figure}

Assuming Z production with a rapidity distribution 
symmetric in the proton-contributor reference frame (see Eq.\ref{Eq:hypo}) 
can provide a phenomenological explanation to the observation made 
by the ATLAS and CMS collaborations.  
As it was quoted above, the rapidity of the proton-contributor reference 
frame for p--Pb ($N_{coll}\approx$6) is -0.430 in the LHC reference frame. 
It is, indeed, observed (see Fig.4 of \cite{ZATLASQM2014:2014}) 
that the measured Z  rapidity distribution exhibits a maximum 
around this value. 
Hence, the Z rapidity distributions for 0\%-10\% ($N_{\rm coll}\approx 14$), 
10\%-40\% ($N_{\rm coll}\approx 10$) and   40\%-90\% ($N_{\rm coll}\approx 4$) 
should be centred at $y^{\rm Z}$ equal to -0.85, -0.69 and -0.23, thus qualitatively 
agreeing with the experimental observations (see Fig.9 of \cite{ZATLASQM2014:2014}).  
At Quark Matter Conference, the CMS collaboration presented the backward-to-forward 
ratio of the Z  rapidity distribution in the proton-nucleon centre of 
mass \cite{Granier:2014, Granier:2014c}.
A Gaussian rapidity distribution with a width equal to 3 
(solid line of Fig.\ref{fig:fig4} top), can be considered to model the 
Powheg-Pythia predictions presented in Fig.2 of reference 
\cite{Granier:2014b, ZCMSQM2014:2014}.  
Assuming that the Z  rapidity distribution in p--Pb collisions 
has the same shape but is centred at the rapidity of  the 
proton-contributor  reference frame ($y_{\rm pC}-y_{\rm pN}$ ), the dashed 
curve plotted in Fig.\ref{fig:fig4} top is obtained. The backward-to-forward 
ratio can then been easily calculated as the ratio of the dashed 
and solid curve in Fig.\ref{fig:fig4} top and it is plotted in Fig.\ref{fig:fig4} bottom. 
This prediction is in agreement with the backward-to-forward ratio measured
by the CMS collaboration \cite{ZCMSQM2014:2014}.

\section{Summary and outlooks}

I have defined the proton-contributor the proton-contributor system in p--A collisions as the system formed by the proton and the nucleons of the nucleus participating to the collisions, which is determined with the Glauber model. Assuming that the particle rapidity distribution is identical to that in pp collisions but centred at the rapidity of the proton-contributor system,
\begin{itemize}
\item the pattern of the pseudo-rapidty distribution ratios of charged particles as a function of the collision centrality in p--Pb at 5.02 TeV, measured by the ATLAS collaboration \cite{ATLASQM2014:2014, DebbeVelasco:2014},
\item the nuclear modification of J/$\psi$ at forward and backward rapidity in p--Pb collisions at 5.02 TeV measured by the ALICE collaboration \cite{Abelev:2013yxa, Lakomov:2014},  and
\item the backward-to-forward ratio of Z bosons in p--Pb collisions at 5.02 TeV measured by the CMS collaboration  \cite{ZCMSQM2014:2014, Granier:2014} 
\end{itemize}
can be understood.

This phenomenological observation might trigger new theoretical ideas on the physics underlying the present hypothesis of the proton-contributor frame.  Seemingly, this hypothesis could be also applied to RHIC results in d-Au collisions and to other observables in p--Pb collisions, like $\Upsilon$ or W production.
Finally, one could imagine new experimental observables taking into account the rapidity gap $\Delta y _{\rm pN-pC}$ between the proton-nucleon and the proton-contributor systems, like the the backward-to-forward ratio in the proton-contributor frame or a proton-contributor nuclear modification factor that would be defined as:
\begin{equation}
R^{pC}_{\rm pA}(y)= \frac{Y_{\rm pA} (y)}{\langle N_{\rm coll} \rangle Y_{\rm pp}(y-\Delta y _{\rm pN-pC})}
\end{equation}

This is a preliminary draft and comments and suggestions are welcome. I plan to make an oral presentation during the 
\href{http://llr.in2p3.fr/sites/qgp2014/index.php}{\emph{Rencontres QGP-France} September 15th-18th 2014 in Etretat}.

\section{acknowledgements}

I would like to thank Philippe Crochet and  St\'ephane Peign\'e for the fruitful and inspiring discussions, Raphael Granier de Cassagnac and Anna Zsigmond for providing the Powheg-Pythia numerical values from \cite{ZCMSQM2014:2014}, and Philippe Crochet and Ombretta Pinazza for the careful reading of the manuscript.

\bibliography{ReferenceInpA}

\begin{thebibliography}{12}
\expandafter\ifx\csname natexlab\endcsname\relax\def\natexlab#1{#1}\fi
\expandafter\ifx\csname bibnamefont\endcsname\relax
  \def\bibnamefont#1{#1}\fi
\expandafter\ifx\csname bibfnamefont\endcsname\relax
  \def\bibfnamefont#1{#1}\fi
\expandafter\ifx\csname citenamefont\endcsname\relax
  \def\citenamefont#1{#1}\fi
\expandafter\ifx\csname url\endcsname\relax
  \def\url#1{\texttt{#1}}\fi
\expandafter\ifx\csname urlprefix\endcsname\relax\def\urlprefix{URL }\fi
\providecommand{\bibinfo}[2]{#2}
\providecommand{\eprint}[2][]{\url{#2}}

\bibitem[{\citenamefont{Aad et~al.}(2013)}]{ATLASQM2014:2014}
\bibinfo{author}{\bibfnamefont{G.}~\bibnamefont{Aad}} \bibnamefont{et~al.}
  (\bibinfo{collaboration}{ATLAS Collaboration}) (\bibinfo{year}{2013}),
  \eprint{\href{https://cds.cern.ch/record/1599773?ln=en}{ATLAS-CONF-2013-096}}.

\bibitem[{\citenamefont{Debbe-Velasco et~al.}(2014)}]{DebbeVelasco:2014}
\bibinfo{author}{\bibfnamefont{R.~R.} \bibnamefont{Debbe-Velasco}}
  \bibnamefont{et~al.} (\bibinfo{collaboration}{ATLAS Collaboration}),
  \bibinfo{journal}{\href{https://indico.cern.ch/event/219436/session/9/contribution/225}{Quark
  Matter Presentation, May 19-24th, Darmstadt}}  (\bibinfo{year}{2014}).

\bibitem[{\citenamefont{Abelev et~al.}(2014)}]{Abelev:2013yxa}
\bibinfo{author}{\bibfnamefont{B.~B.} \bibnamefont{Abelev}}
  \bibnamefont{et~al.} (\bibinfo{collaboration}{ALICE Collaboration}),
  \bibinfo{journal}{JHEP} \textbf{\bibinfo{volume}{1402}}, \bibinfo{pages}{073}
  (\bibinfo{year}{2014}), \eprint{1308.6726}.

\bibitem[{\citenamefont{Bossu et~al.}(2011)\citenamefont{Bossu, Conesa~del
  Valle, de~Falco, Gagliardi, Grigoryan et~al.}}]{Bossu:2011qe}
\bibinfo{author}{\bibfnamefont{F.}~\bibnamefont{Bossu}},
  \bibinfo{author}{\bibfnamefont{Z.}~\bibnamefont{Conesa~del Valle}},
  \bibinfo{author}{\bibfnamefont{A.}~\bibnamefont{de~Falco}},
  \bibinfo{author}{\bibfnamefont{M.}~\bibnamefont{Gagliardi}},
  \bibinfo{author}{\bibfnamefont{S.}~\bibnamefont{Grigoryan}},
  \bibnamefont{et~al.} (\bibinfo{year}{2011}), \eprint{1103.2394}.

\bibitem[{\citenamefont{Mart\'{\i}n-Blanco et~al.}(2014)}]{MartinBlanco:2014}
\bibinfo{author}{\bibfnamefont{J.}~\bibnamefont{Mart\'{\i}n-Blanco}}
  \bibnamefont{et~al.} (\bibinfo{collaboration}{ALICE Collaboration}),
  \bibinfo{journal}{\href{https://indico.cern.ch/event/219436/session/17/contribution/135}{Quark
  Matter Presentation, May 19-24th, Darmstadt}}  (\bibinfo{year}{2014}).

\bibitem[{\citenamefont{Lakomov et~al.}(2014)}]{Lakomov:2014}
\bibinfo{author}{\bibfnamefont{I.}~\bibnamefont{Lakomov}} \bibnamefont{et~al.}
  (\bibinfo{collaboration}{ALICE Collaboration}),
  \bibinfo{journal}{\href{https://indico.cern.ch/event/219436/session/2/contribution/129}{Quark
  Matter Presentation, May 19-24th, Darmstadt}}  (\bibinfo{year}{2014}).

\bibitem[{\citenamefont{Aad et~al.}(2014)}]{ZATLASQM2014:2014}
\bibinfo{author}{\bibfnamefont{G.}~\bibnamefont{Aad}} \bibnamefont{et~al.}
  (\bibinfo{collaboration}{ATLAS Collaboration}) (\bibinfo{year}{2014}),
  \eprint{\href{http://cds.cern.ch/record/1702971}{ATLAS-CONF-2014-020}}.

\bibitem[{\citenamefont{Chatrchyan et~al.}(2014)}]{ZCMSQM2014:2014}
\bibinfo{author}{\bibfnamefont{S.}~\bibnamefont{Chatrchyan}}
  \bibnamefont{et~al.} (\bibinfo{collaboration}{CMS Collaboration})
  (\bibinfo{year}{2014}),
  \eprint{\href{http://cms-physics.web.cern.ch/cms-physics/public/HIN-14-003-pas.pdf}{HIN-14-003}}.

\bibitem[{\citenamefont{Cole et~al.}(2014)}]{Cole:2014}
\bibinfo{author}{\bibfnamefont{B.}~\bibnamefont{Cole}} \bibnamefont{et~al.}
  (\bibinfo{collaboration}{ATLAS Collaboration}),
  \bibinfo{journal}{\href{https://indico.cern.ch/event/219436/session/3/contribution/715}{Quark
  Matter Presentation, May 19-24th, Darmstadt}}  (\bibinfo{year}{2014}).

\bibitem[{\citenamefont{de~Cassagnac
  et~al.}(2014{\natexlab{a}})}]{Granier:2014}
\bibinfo{author}{\bibfnamefont{R.~G.} \bibnamefont{de~Cassagnac}}
  \bibnamefont{et~al.} (\bibinfo{collaboration}{CMS Collaboration}),
  \bibinfo{journal}{\href{https://indico.cern.ch/event/219436/session/3/contribution/712}{Quark
  Matter Presentation, May 19-24th, Darmstadt}}
  (\bibinfo{year}{2014}{\natexlab{a}}).

\bibitem[{\citenamefont{de~Cassagnac
  et~al.}(2014{\natexlab{b}})}]{Granier:2014c}
\bibinfo{author}{\bibfnamefont{R.~G.} \bibnamefont{de~Cassagnac}}
  \bibnamefont{et~al.} (\bibinfo{collaboration}{CMS Collaboration}),
  \bibinfo{journal}{\href{https://cds.cern.ch/record/1747544?ln=fr}{Quark
  Matter Proceedings, May 19-24th, Darmstadt}}
  (\bibinfo{year}{2014}{\natexlab{b}}).

\bibitem[{\citenamefont{de~Cassagnac and Zsigmond}(2014)}]{Granier:2014b}
\bibinfo{author}{\bibfnamefont{R.~G.} \bibnamefont{de~Cassagnac}}
  \bibnamefont{and} \bibinfo{author}{\bibfnamefont{A.}~\bibnamefont{Zsigmond}},
  \bibinfo{journal}{Private communication}  (\bibinfo{year}{2014}).

\end{thebibliography}

\end{document}